\documentclass[letterpaper, 10 pt, conference]{ieeeconf}  

\IEEEoverridecommandlockouts                              
\title{\LARGE \bf
Process Mining of Programmable Logic Controllers: Input/Output Event Logs
}

\usepackage{graphicx}
\usepackage{xcolor}
\let\labelindent\relax
\usepackage{enumitem}
\usepackage{algorithm}
\usepackage[noend]{algpseudocode}
\usepackage{amsmath}

\makeatletter
\renewcommand{\ALG@beginalgorithmic}{\small}
\makeatother

\author{{Julian~Theis,~\IEEEmembership{Graduate Student Member,~IEEE}, Ilia~Mokhtarian,
and~Houshang~Darabi,~\IEEEmembership{Senior Member,~IEEE}}
\thanks{J. Theis, I. Mokhtarian, and H. Darabi are with University of Illinois at Chicago, Department of Mechanical and Industrial Engineering, 842 West Taylor Street, Chicago, IL 60607, United States. H. Darabi is the corresponding author. E-mail: \{jtheis3, imokht2, hdarabi\}@uic.edu.}}

\begin{document}

\maketitle
\thispagestyle{empty}
\pagestyle{empty}
\begin{abstract}
This paper presents an approach to model an unknown Ladder Logic based Programmable Logic Controller (PLC) program consisting of Boolean logic and counters using Process Mining techniques. First, we tap the inputs and outputs of a PLC to create a data flow log. Second, we propose a method to translate the obtained data flow log to an event log suitable for Process Mining. In a third step, we propose a hybrid Petri net (PN) and neural network approach to approximate the logic of the actual underlying PLC program. We demonstrate the applicability of our proposed approach on a case study with three simulated scenarios. 


\end{abstract}

\section{Introduction}\label{sec:introduction}
Programmable Logic Controllers (PLCs) have a long-standing history and are omnipresent in industrial control applications \cite{darabi2002ladder} on a global scale. First introduced in the 1960s, these controllers have survived technological and workforce generation changes without major conceptual modifications. Due to their ease of use, reliability, and wide adoption, PLCs are likely to coexist and to integrate with technological advancements such as internet-connected sensors and advanced data systems, described under the terms \textit{Industrial Internet of Things} (IIoT) and \textit{Industry 4.0} \cite{iiot1}. However, the aging of PLC programs is increasingly becoming a problem for organizations, mainly due to the departure of workforce and anachronistic documentation guidelines \cite{lacofdocumentation}. In the worst case, this means that the logic of long-standing existing programs is no longer understood or access to program source codes is completely lost. Extending or debugging PLCs to adopt new technologies is non-trivial and carries many dangers when exposing process components to the internet through IIoT \cite{PLCIoTthreat}.



In this paper, we propose an approach which unveils the logic of an unknown, i.e. black box PLC program consisting of Boolean and counter components using Process Mining techniques to overcome the real-world problems introduced above. We first tap the inputs and outputs of the PLC during runtime and record all data flows over time. Then, we leverage the recordings to convert them to event logs which are suitable for Process Mining. We finally approximate the black box PLC program by discovering a process model and extending it using an existing state-of-the-art event prediction algorithm called \textit{DREAM-NAP} \cite{dream-nap}. We derive a set of rules in order to deploy our approximated controller and to replace the black box. The applicability of this approach is demonstrated on a case study with three simulated scenarios. 


This paper is structured as follows. Section \ref{sec:related-work} discusses related work. We introduce required preliminaries in Section \ref{sec:preliminaries} followed by a formal problem definition in Section \ref{sec:problemdef}. Our approach is proposed in Section \ref{sec:approach} which is evaluated using a case study in Section \ref{sec:casestudy}. We conclude this paper and discuss further research aspirations in Section \ref{sec:conclusion}.


\section{Related Work}\label{sec:related-work}
Most of the existing PLCs are programmed in Ladder Logic Diagram (LLD) \cite{darabi2002ladder}. In many cases, applied LLDs are large, non-structured programs which are difficult to debug and maintain \cite{abdelhameed2005diagnosis, darabi2009neural}, \cite{darabi2014design}. However, LLD has the advantage of being convertable to Petri nets (PNs). Therefore, the research community has developed several approaches for the translation of Ladder Logic to PNs in order to examine the behavior of a controller.

A first overview of early methods converting from LLD and Relay Logic to PNs can be found in the work of Peng and Zhou \cite{peng2004ladder}. It includes the approach proposed by Lee and Lee \cite{bum2002conversion} to convert an existing LLD to a PN using the controller characteristics such that one can formally analyze LLD programs. 

Similarly, Lee and Lee \cite{lee2009conversion} convert LLDs to PNs using the so-called modulus synthesis technique. This approach first translates the basic LLD components to PN elements, followed by complex structures such as counters and timers.

Da Silva Oliveira et al. \cite{da2011obtaining} proposed a conversion algorithm to formally verify LLD programs using Colored PNs. This method is capable of modeling counters and timers.

Chen et al. \cite{chen2012method} propose an algorithm to convert LLDs to ordinary PNs to validate and debug controller applications. They introduce a systematic approach which demonstrates advantages over element-wise conversion. However, this work converts only basic LLD components and disregards complex structures like counters and timers. 

Quezada et al. \cite{Quezada2014} developed a further element-wise conversion method of LLD to PNs. Specifically, the authors translate five predefined popular control line types to a PN which they call a \textit{Ladder Diagram Petri Net}. Quezada et al. demonstrate the applicability of their method on two different case studies. However, this approach does not consider counters and timers and is limited to the predefined control structures.

Luo et al. \cite{luo2018modeling} introduce an approach with two major contributions towards the translation of LLDs to PNs. First, they translate an LLD to an ordinary PN rather than extending a PN. Second, the authors develop a systematic approach to translate an LLD to PN at once.


The previous literature is based on the assumption that LLD source codes are given. Moreover, only some of the methods translate non-Boolean LLD structures to PNs. We therefore propose a data-driven approach to convert an unknown LLD which encompasses counters to a hybrid PN which mimics the actual underlying controller program.


\section{Preliminaries}\label{sec:preliminaries}
This section introduces preliminaries which are required throughout this paper. We specifically introduce Ladder Logic, 
Event Logs, Petri Nets, Process Mining, a process discovery algorithm called \textit{Split Miner}, and a next activitiy prediction algorithm called \textit{DREAM-NAP}.

\subsection{Ladder Logic}\label{sec:prelim-ladderlogic}
Ladder Logic is a graphical programming language which is similar to the schematic of relay control circuits. It is the most used programming language to develop PLC applications. In general, an LLD consists of two vertical lines, called rails, and a set of horizontal lines, called rungs. Additionally, such a program encompasses multiple binary and continues variables which can be defined as either inputs, outputs, or internal variables. A PLC executes all rungs sequentially from top to bottom and from the left to the right rail. A rung can be associated with several logical operations to modify variables. The set of logical operators include simple Boolean components such as XORs as well as more complex structures like counters and timers.

\subsection{Event Logs}\label{sec:prelim-eventlogs}
The subsequent definitions are based on the work of van der Aalst \cite{processmining}.
An event can be any real-world observable action and is described by an activity $u$ and a timestamp $t$. Optionally, an event may encompass further attributes such as associated resources, costs, or people. We distinguish between two events using unique identifiers $e$. We define $E$ as the set of all possible event identifiers and $U$ as the set of all possible activities. The set of all possible attributes is denoted by $D$. It follows that for each $e \in E$ and any $d \in D: \#_d(e)$ is the value of the attribute $d$ of the event identified by $e$. In the case that an event $e$ does not contain an attribute $d$, the value will be null, i.e. $ \#_d(e) = \perp$. We define a further function $\gamma : E \rightarrow A$ which maps each event to an activity.


A trace $k \in K$ is defined as a finite sequence of events belonging to the same process run where $K$ is the set of all possible traces. Each trace consists of at least of one event and events within a trace are ordered chronologically.

An Event Log $L$ is a finite set of traces such that $L \subseteq K$. Each event occurs only once in the entire log.

\subsection{Petri Net}\label{sec:prelim-petrinets}
A PN is a tool which is used for process modeling in many different areas such as healthcare \cite{healthcare}, project management \cite{darabirev2}, and manufacturing \cite{manufacturing}. A PN can be defined as a quadruplet $PN = (P, T, F, \pi)$ where $P$ is a set of places, $T$ is a set of transitions, $F \subseteq (P \times T) \cup (T \times P)$ is a set of directed arcs, and $\pi$ is a function that maps transitions to activities \cite{vanderAalst2016}. When visualizing PNs, places are represented as circles whereas transitions are represented as rectangles. The set of $F$ is visualized as unidirectional arcs connecting transitions to places and vice versa. The function $\pi : T \rightarrow U \cup \{\perp\}$ maps each transition to either an activity $u \in U$ or to a non-observable activity. Transitions which map to non-observable activities are called \textit{hidden transitions} and are visualized by a black rectangle.

Each place of a PN can hold a non-negative integer number of tokens. The distribution of tokens in a PN during runtime is usually represented in a vector format and is called the marking $M$ of a PN. Each PN has at least one initial and one final marking corresponding to the start and the end of a process run. 

During runtime, a transition can fire if all input places of the same transition hold at least one token. We call this transition \textit{enabled}. It fires when a corresponding event activity is observed. The number of tokens of each input place of a fired transition is reduced by $1$, whereas the number of tokens at each output place of the fired transition is increased by $1$. 


\subsection{Process Mining}\label{sec:prelim-processmining}
Process Mining describes the three subcategories of process discovery, conformance checking, and enhancement of process models \cite{vanderAalst2016}. Process discovery represents techniques to obtain process models from event logs. With conformance checking, one is particularly interested in the evaluation of the quality of an obtained process model, i.e. how well a model represents a process. Enhancement describes the improvement of existing process models by considering additional information such as performance data.


\subsection{Split Miner}\label{sec:prelim-splitminer}
\textit{Split Miner} \cite{splitminer} is a recent PN process model discovery algorithm which demonstrated major performance improvements compared to previous methods \cite{pnbenchmark}. This approach has been developed to tackle the tradeoff between several process conformance measures.

\subsection{DREAM-NAP}\label{sec:prelim-decayfc}
\textit{DREAM-NAP} is a state-of-the-art method to predict next activities in running process cases and stands for \textit{Decay Replay Mining - Next Activity Prediction} \cite{dream-nap}. This method extends a PN process model with decay functions associated to each place. During replay, the activation function of a specific place is activated as soon as a token enters, and decays over time until a new token enters the same place. Moreover, token movements in each place are counted over time. Using these mechanisms, \textit{DREAM-NAP} creates timed PN state samples consisting of a decay value vector of all places, token movement counters, and the current marking of the PN while replaying an event log. These samples are used to train a fully connected neural network to predict the next activity of a running process. In this way, possible correlations between time intervals of occurring events are used to forecast upcoming activities in a more flexible way than by e.g. deadline-based time PNs \cite{darabirev1, darabirev3}.
DREAM-NAP demonstrated significant performance improvements over existing state-of-the-art methods on a diverse set of benchmark datasets.


\section{Problem Definition}\label{sec:problemdef}
The research community developed different white box modeling techniques to convert controller logic to process models such as PNs. Recent methods can easily translate Boolean logic to desired models. However, these approaches face issues when the logic comprises complex structures such as continuous variables, delays, and/or counters. These problems intensify when the logic is unknown and inaccessible, i.e. when being confronted with a black box rather than a white box. In this case, one can only gain access to the physical inputs and outputs of the controller. 

Lets assume we are given a controller $C$ which consists of a set of physical binary inputs $I$, a set of physical binary outputs $O$, and a control logic program $\delta$. The controller $C$ controls a process $P$ which in turn consists of a set of binary sensors $S$ and a set of binary actuators $A$. Each sensor $s \in S$ is connected to one or many inputs $i \in I$, whereas each output $o \in O$ is connected to one or many actuators $a \in A$. Each input $i \in I$ and each output $o \in O$ can either be $0$ or $1$ at any time $t$. We denote the value of an input and output at time $t$ as $i(t)$ and $o(t)$ respectively. A control logic program $\delta(t): I(t) \rightarrow O(t)$ maps the set of input values to a set of output values at time $t$. The program's logic $\delta(t)$ encompasses Boolean logic and optionally counters. The control logic program, $\delta(t)$, is unknown, therefore one is confronted with a black box problem.

Consequently, the objective is to detect a controller model $C'$ which consists of the same sets of inputs and outputs like the actual controller $C$ and which approximates the complex control logic $\delta$ denoted by $\delta '$. At its underlying core, the approximated model $C'$ is required to be interpretable such that $C'$ can be visually inspected and debugged effortlessly.

\section{Approach}\label{sec:approach}
\begin{figure}[h]
  \begin{center}
    \includegraphics[width=250pt]{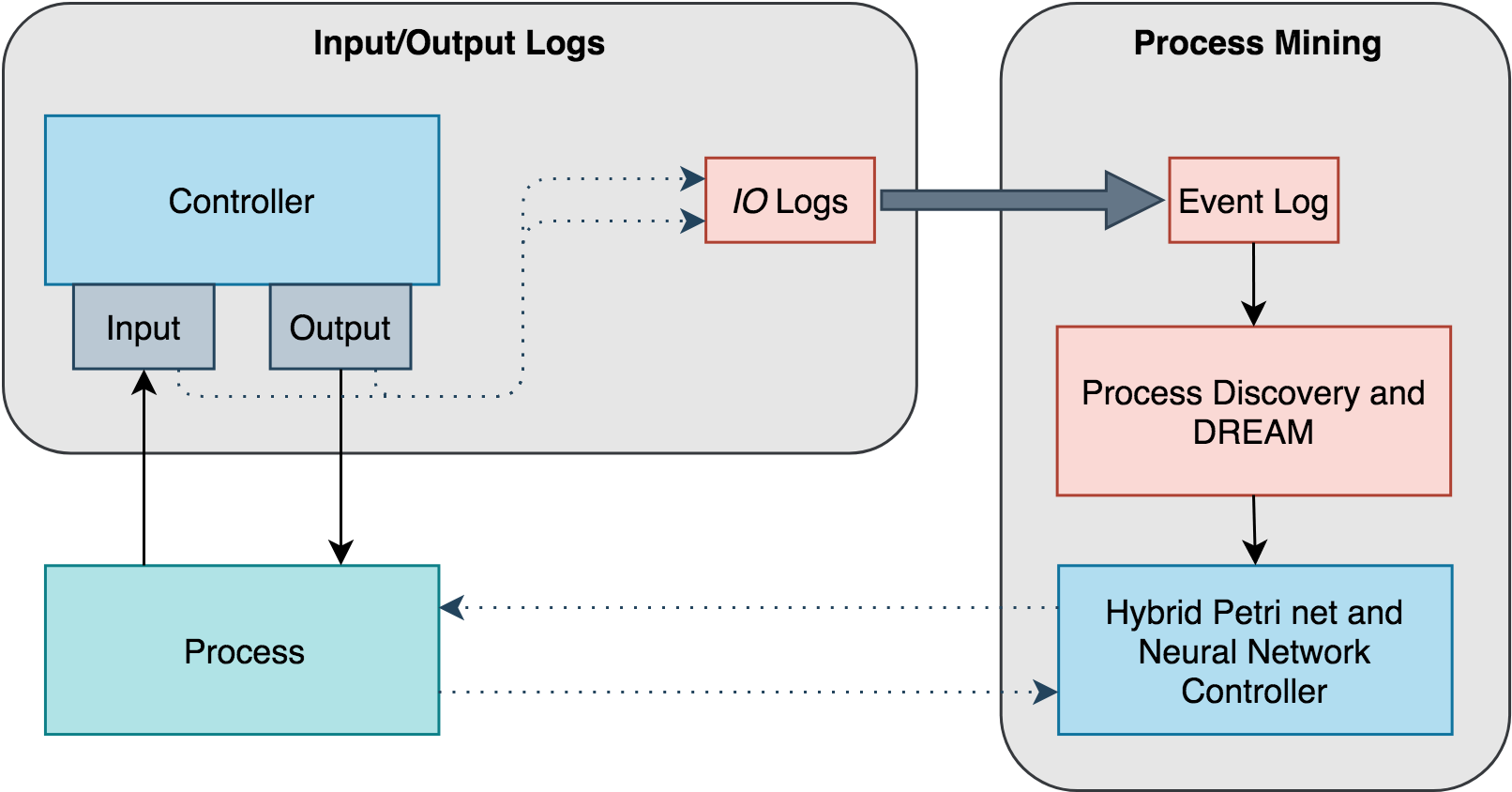}
  \end{center}
  \caption{Overview of the proposed approach which consists of the two main blocks: Input/Output Logs and Process Mining.}
  \label{fig:approach}
\end{figure}

In this section, we propose an approach to discover an interpretable approximate controller $C'$ from a black box controller $C$ which contains Boolean logic and counters. We introduce a hybrid PN and neural network approach which unveils the logic of a control program. First, we tap the binary inputs and outputs of a controller and record the data flows over time. In a second step, we take these recordings and convert them to an event log with multiple traces which is suitable for Process Mining purposes. From this event log, we discover a process model and define an approximate controller $C'$. The overview of the proposed approach is visualized in Figure \ref{fig:approach}.

\subsection{Input/Output Logs}
Initially, we tap the input and output modules of a PLC $C$ to record the data flows. These recordings are stored in a raw Input/Output log denoted by $IO$. We observe all input and output values over time $t$ at every PLC scan cycle and write the input values $I(t)$ followed by the output values $O(t)$ interleaved and in chronological order to $IO$. Naturally, $O(t)$ has to follow $I(t)$ since the controller program $\delta(t)$ sets output values based on input values. In this way, we create a log file in which each row represents a sample consisting of a timestamp $d(t)$ corresponding to the read time $t$, the input/output address identifying a specific $i$ or $o$, and the corresponding recorded value $i(t)$ or $o(t)$. 

All elements of $I$ and $O$ must be tapped for a sufficient amount of time such that the Input/Output log $IO$ embodies a representative sample of the behavior of the controller program $\delta$.
The recording duration is highly dependent on the size of the controlled process, the control logic program complexity, and the number of distinct performed control actions on a process $P$.

\subsection{Process Mining}
To apply Process Mining techniques, we must convert the Input/Output log $IO$ to an event log $L$ which satisfies the definitions of Section \ref{sec:prelim-eventlogs}.
This means that we have to translate the obtained samples to events. Earlier, we defined an event as any observable real-world activity. In our context, an activity describes a change of a specific input or output value, i.e. a change from $0$ to $1$ or vice versa, at time $t$. Thus, an event activity comprises the address of the input or output and the new value. Each event carries additionally an attribute called \textit{class} with two possible values, $\%I$ and $\%Q$, to label the physical address as either \textit{input} or \textit{output}.
The timestamp remains unchanged.

We parse the Input/Output log $IO$ such that we obtain a reduced log $IO_{r}$ with samples described in the format above. Each event in $IO_r$ reflects a change of value of an input or output at timestamp $d(t)$ compared to time $t-1$. We further compress $IO_r$ by merging all events of class $\%I$ with identical $d(t)$ into one event  where the event's activity is the concatenation of activities of the merged events. Similarly, we merge all events of class $\%Q$ with identical $d(t)$. 

Next, we convert $IO_r$ to an event log $L$ by splitting the sequence of events in $IO_r$ into a set of multiple traces, as required per definition in Section \ref{sec:prelim-eventlogs}. We slice the raw log into multiple traces based on the occurrence of a \textit{reset activity} $r$. The reset $r$ is an activity which is easily recognizable as such from a process perspective and which resets the process to its initial state. We assume that such an activity exists and is known. The method is illustrated in Algorithm \ref{algo:ior}.

\begin{algorithm}
    \caption{$IO$ to $L$}\label{algo:ior}
    \begin{algorithmic}[1]
       \State $\textit{activity} \gets \textit{null},~\textit{cls} \gets \textit{IO.0.cls},~\textit{t} \gets 
       \textit{IO.0.t}$
       \State $\textit{IOr} \gets [~], \textit{L} \gets \{~\}$, $\textit{trace} \gets [~]$
       
       \For{$io$ in $IO$}
        \If {$tChange(io.t, t)$ or $clsChange(io.cls, cls)$} 
            \State $\textit{IOr}.append(activity)$
            \State $\textit{activity} \gets \textit{io.activity}$
        \Else
            \State $\textit{activity}.append(io.activity)$
            \State $\textit{cls} \gets io.cls$
            \State $\textit{t} \gets io.t$
        \EndIf
        
       \EndFor
       \State $\textit{IOr}.append(activity)$
       
       \For{$io$ in $IOr$}
        \If {$io$~contains~$r$} 
            \State $\textit{L}.append(trace)$
            \State $\textit{trace} \gets [~]$
        \EndIf
        \State $\textit{trace}.append(io)$
       \EndFor
       \State $\textit{L}.append(trace)$
    \end{algorithmic}
\end{algorithm}

Finally, \textit{Split Miner} is used to discover a PN from the event log $L$. We extend the obtained process model with a function $\gamma$ mapping each transition to either class $\%I$, class $\%Q$, or $\perp$ (null). All transitions of class $\%I$ correspond to PLC input activities, i.e. data communicated from a process. Similarly, transitions of class $\%Q$ correspond to PLC output activities, i.e. actuator settings communicated from the PLC to a process. Transitions which do not belong to either class, thus $\perp$, are hidden transitions and are required to model the PLCs logic using PNs.

We use $\gamma$ to derive the subsequent rules when defining an approximate controller $C '$ as a substitution of a true controller $C$:
\begin{enumerate}
  \item Whenever there are only $\%I$ class transitions enabled, the approximate controller $C '$ has to wait until one of the corresponding activities of these transitions will occur.
  \item Whenever there is only one $\%Q$ class transition enabled, this transition will fire immediately and the corresponding approximate controller output will bet set correspondingly.
  \item Whenever there are multiple class $\%Q$ transitions enabled, or whenever there is at least one class $\%I$ and one class $\%Q$ transition enabled, the approximate controller $C '$ has to decide to either await specific input values or to set specific output values.
\end{enumerate}
Where Rules 1 and 2 are straightforward, Rule 3 requires an additional functionality introduced to the approximate controller $C '$. We leverage a time-aware method called \textit{DREAM-NAP} which has been introduced in Section \ref{sec:prelim-decayfc} \cite{dream-nap}. This method extends every place in the PN with a time decay function and introduces token movement counters in order to create timed PN state samples during replay or runtime. These samples are used to train a neural network which accurately predicts the next event activity. Since we can classify the next activity using $\gamma$, the approximate controller $C '$ can react accordingly by either awaiting a predicted input value or by setting specific output values. The approximate controller $C'$ logic is described in Algorithm \ref{algo:control}.
\begin{algorithm}
    \caption{$C'$ controller}\label{algo:control}
    \begin{algorithmic}
       \While{$C'~running$}
        \State $\textit{trans} \gets PN.enabledTransitions()$
        \If{$trans.size() < 2 \And trans.contains(\%Q)$} 
            \State $executeOutput(trans)$
        \Else
            \If{$!trans.contains(\%Q)$} 
                \State $waitForInputEvent(trans)$
            \Else
                 \State $op \gets \textit{DREAM-NAP(PN)}$
                 \State $executeOrWait(op)$
            \EndIf
        \EndIf
       \EndWhile
    \end{algorithmic}
\end{algorithm}

\section{Case Study: Tank Level Control}\label{sec:casestudy}
We demonstrate the applicability of our proposed approach by providing a case study of a tank level control system which can be found in different industrial applications such as liquids filtration, food processing, or water purification plants \cite{casestudy1, casestudy2}.
We developed an experimental environment consisting of a simulated plant environment and a LLD based control program running on top of \textit{OpenPLC}, an open source PLC software \cite{openplc}. 

\subsection{Tank Level Control Simulation}
At the core of the plant is a tank with a maximum volume of $100$ gallons. The tank possesses an inward flow and an outward flow. The inward flow is controlled by a binary valve denoted by $inv$, i.e. it is either completely opened or completely closed, respectively $1$ or $0$. The outward flow is controlled accordingly with a binary outward valve denoted by $outv$, in which $1$ represents the opened and $0$ the closed valve. It follows that $\{inv,outv\}$ is the set of actuators $A$ of the plant.

The set of sensors $S$ of this plant consists of an upper-level sensor $ULS$ at tank level $90$, a middle level sensor $MLS$ at tank level $50$, and a lower-level sensors $LLS$ at tank level $10$, $S = \{ULS, MLS, LLS\}$. Each sensor reports either $1$ or $0$ depending if the corresponding sensor detects liquid at its location inside the tank.

We introduce two versions of this plant, $P_1$ and $P_2$. In plant setup $P_1$, we do not consider any uncertainty, i.e. the inward and outward flows are set to a constant rate of $9$ gallons per second. Furthermore, a closed valve sets the corresponding flow to $0$ gallons per second. 
In contrast, $P_2$ draws the inward and outward flow rates from a normal distribution with mean $9$ and standard deviation $2$ gallons per second. In case of a closed valve, we allow for some leakage drawn from a uniform distribution with lower limit $0$ and upper limit $0.5$ gallons per second.

\subsection{PLC Programs}
We define two LLD based PLC programs to control the plants setups described above.
\begin{figure}[!h]
  \begin{center}
    \includegraphics[width=170pt]{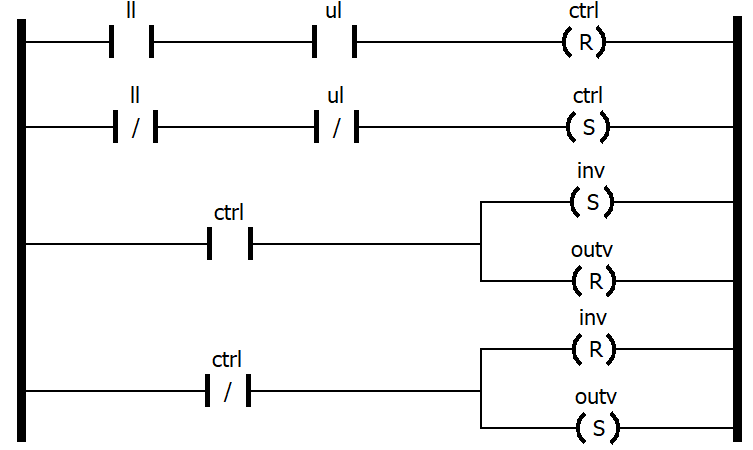}
    \caption{LLD of the $C_1$ PLC program. The sensor variables are denoted by \textit{ll} and \textit{ul} corresponding to $LLS$ and $ULS$ respectively. The actuators of $inv$ and $outv$ are set based on a control variable \textit{ctrl}.}
  \end{center}
  \label{fig:c1}
\end{figure}
The first PLC program controls the tank level such that it fills up to the threshold of the $ULS$ sensor before it decreases the level to the threshold of the $LLS$ sensor threshold. We denote this program by $C_1$. Figure 2 visualizes the LLD of this control program.

\begin{figure}[!h]
  \begin{center}
    \includegraphics[width=170pt]{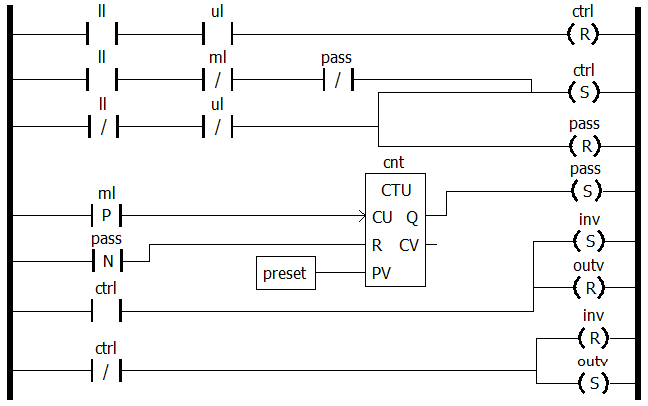}
    \caption{LLD of the $C_2$ PLC program. The sensor variables are denoted by \textit{ll}, \textit{ml}, and \textit{ul} corresponding to $LLS$, $MLS$, and $ULS$ respectively. The variable \textit{ctrl} is used to control the actuators. We introduce a Boolean variable \textit{pass} signaling to drop the tank level below the \textit{LLS} threshold when set to \textit{true}. The count module \textit{cnt} counts from $0$ to \textit{preset = $3$}.}
  \end{center}
  \label{fig:c2}
\end{figure}
The second PLC program is denoted by $C_2$. Initially, this LLD controls the plant such that the tank will be filled up to the $ULS$ threshold. Then, the tank level will be periodically decreased to the $MLS$ and increased to the $ULS$ sensor threshold. As soon as as predefined number of filling and emptying iterations is reached, the level will be decreased below the $LLS$ threshold. The respective LLD is visualized in Figure 3.

Moreover, the sensor variables \textit{ll}, \textit{ml}, and \textit{ul} are mapped to the OpenPLC addresses \textit{\%IX0.1}, \textit{\%IX0.2}, and \textit{\%IX0.0} respectively. The actuators $inv$ and $outv$ are mapped to the addresses \textit{\%QX0.0} and \textit{\%QX0.1} respectively.


\subsection{Experimental Evaluation}
We apply the proposed approach on the following three scenarios:
\begin{itemize}
    \item Scenario 1: $P_1$ controlled by $C_1$
    \item Scenario 2: $P_1$ controlled by $C_2$
    \item Scenario 3: $P_2$ controlled by $C_2$
\end{itemize}
We record Input/Output logs for a duration of $880$ seconds for each of the above scenarios. The reset activity $r$ is defined for all scenarios as \textit{\%IX0.1 false} meaning that one can see that the $LLS$ sensor is exposed in the tank, i.e. the tank is almost empty.

\begin{figure}[!ht]
  \begin{center}
    \includegraphics[width=240pt]{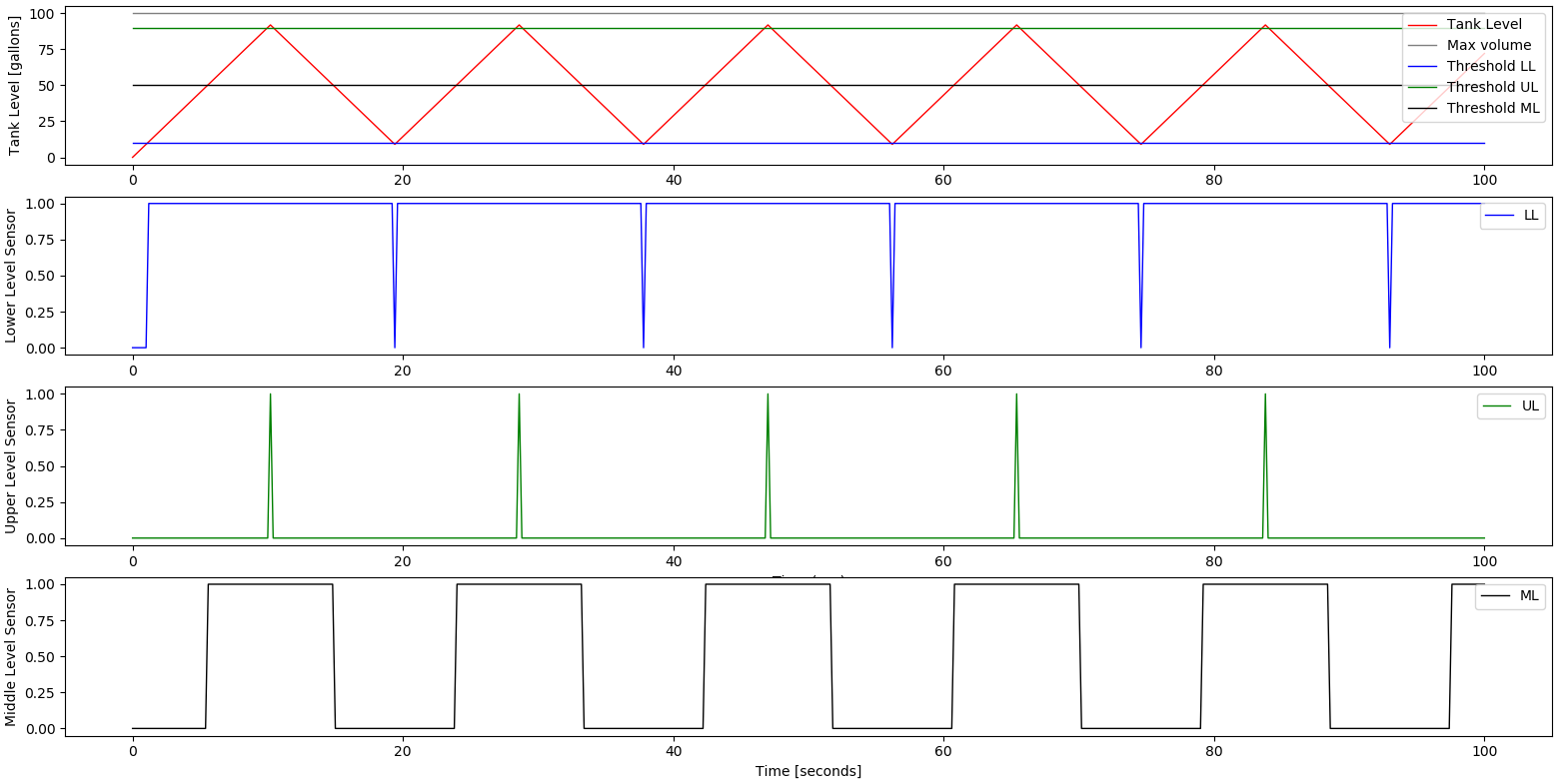}
     \caption{Visualization of Scenario 1. The graph on the top shows the tank level over time. The subsequent graphs visualize the sensor values of $LLS$, $ULS$, and $MLS$.}
  \end{center}
  \label{fig:results_scen1}
\end{figure}

The runtime performance of Scenario 1 is visualized in Figure 4. After $880$ seconds of simulating, we obtain 48 cases to discover an approximate controller $C'$. In this scenario, $C'$ consists of a PN only and is visualized in Figure 5. The PN can be used to control the plant exactly like the LLD PLC program. \textit{DREAM-NAP} is not required since at any time, there is only one $\%Q$ class or one $\%I$ class transition enabled.  

\begin{figure}[!h]
  \begin{center}
    \includegraphics[width=240pt]{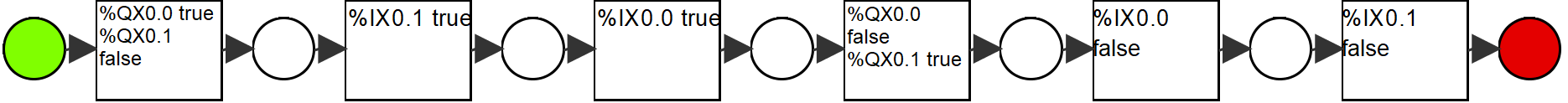}
  \caption{The discovered PN of \textit{Scenario 1} shows a straightforward behavior. The green place represents the initial marking whereas the red place corresponds to the final marking.}
  \end{center}
  \label{fig:results_scen1pn}
\end{figure}

The results of Scenario 2 and 3 can be evaluated together. In both cases, we obtain 22 traces after running the simulation for $880$ seconds. Figure 6 
shows an excerpt of the runtime performance of Scenario 2. When applying our proposed approach, we obtain an identical PN for Scenario 2 and 3. Its structure is visualized in Figure 7. In difference to Scenario 1, this PN allows loops. However, the PN structure itself does not count the number of transition firings. Therefore, we leverage \textit{DREAM-NAP}. We split the event log $L$ into a training set of 17 traces and a testing set of 5 traces. After five training epochs, \textit{DREAM-NAP} predicts next activities with a 100\% accuracy on the test set. We can substitute the LLD PLC programs of Scenario 2 and 3 by applying a combination of the discovered PN with the obtained \textit{DREAM-NAP} neural network model using Algorithm \ref{algo:control}.
\begin{figure}[!h]
  \begin{center}
    \includegraphics[width=247pt]{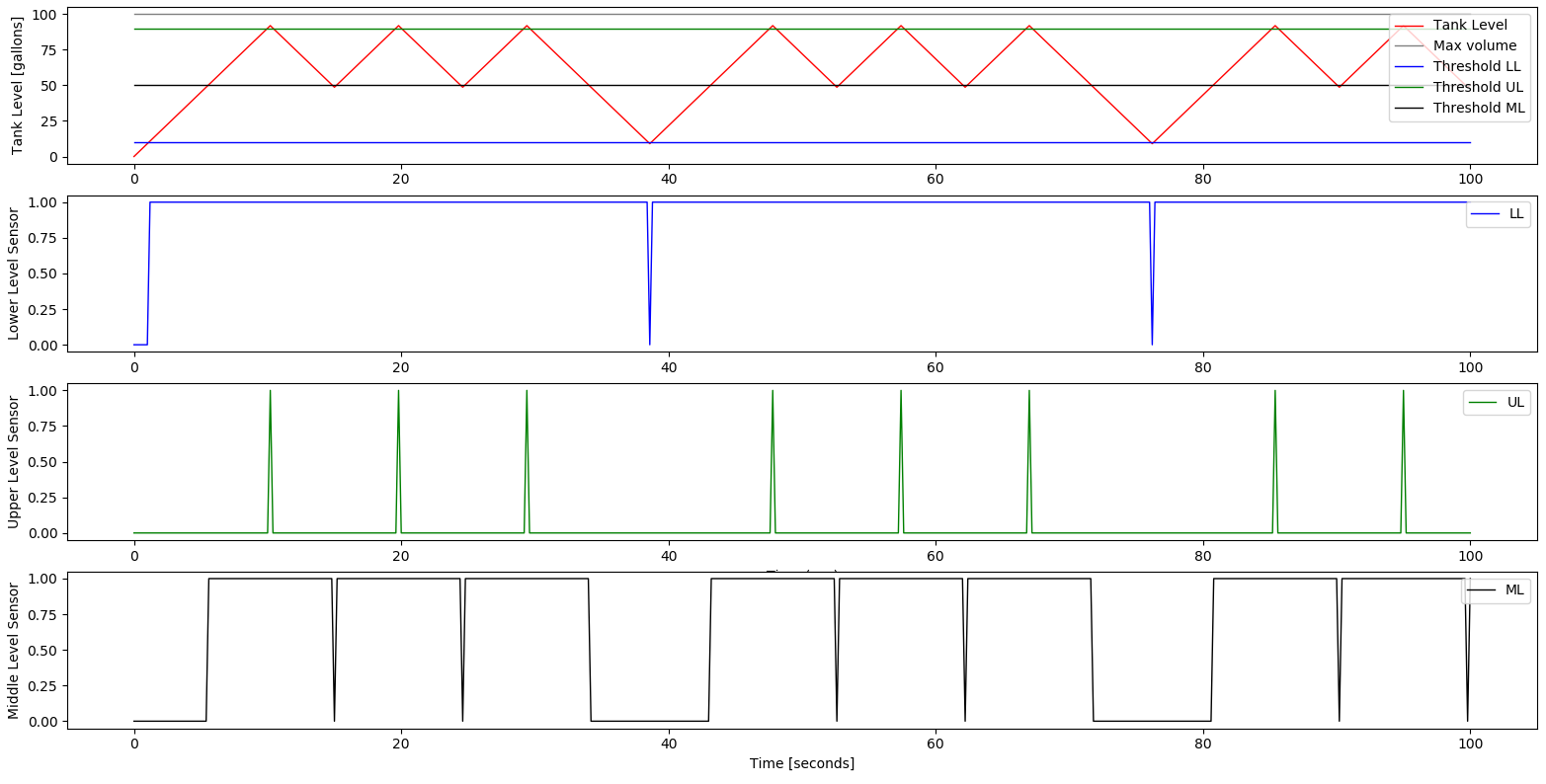}
    \caption{Visualization of the runtime performance of Scenario 2. The graph on the top shows the tank level over time. The subsequent graphs visualize the sensor values of $LLS$, $ULS$, and $MLS$ recorded over time.}
  \end{center}
  \label{fig:results_scen2}
\end{figure}


\begin{figure}[!h]
  \begin{center}
    \includegraphics[width=240pt]{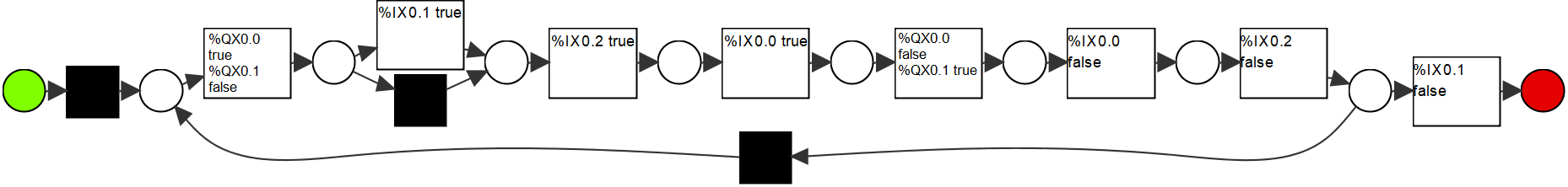}
  \caption{The discovered PN of \textit{Scenario 2} and \textit{Scenario 3} is identical and shows a structure which allows loops. Green places represent the initial marking whereas red places correspond to the final marking.}
  \end{center}
  \label{fig:results_scen2pn}
\end{figure}

\section{Discussion and Conclusion}\label{sec:conclusion}
In this paper, we proposed an approach to unveil the logic of black box LLD PLC programs which consist of Boolean logic and optionally counter modules. This approach specifically taps the inputs and outputs of a PLC, converts the recorded data flows to an event log, and discovers a hybrid PN and \textit{DREAM-NAP} model. We successfully demonstrated the applicability on a case study with three simulated scenarios. To the best of our knowledge, we are the first ones translating an unknown LLD program consisting of complex structures such as counters to a process model by considering only the inputs and outputs of a controller.

Based on the results of our work, we encourage to conduct research studies on the evaluation of the derived rules on further real-world processes and complex controller program structures beyond the introduced case study scenarios. In particular, we encourage the assessment of our approach by considering multiple dependent and independent counters in controller programs. Moreover, further research should be conducted to model timer components and controller-plant setups encompassing continuous variables inputs and outputs to extend the proposed method towards an exhaustive approach. Ideally, such an approach should be capable of modeling all complex structures which are defined in the LLD standard.

Future research and ongoing advancements will enable a wide range of impactful applications.
Since it is crucial to unveil risks in industrial applications, one can leverage our approach to detect safety threats by comprehensively analyzing the obtained process models. 
Additionally, one can easily deploy the models to perform thorough simulation analysis. In this way, organizations can test their setups for robustness and assess disaster behavior.
Furthermore, one can consider a PLC program as a white box and apply our approach. Disclosing logical misalignments between an approximate and the actual known controller can be used to draw inferences about misconfigurations and programming errors.

\addtolength{\textheight}{-12cm}   
\end{document}